\documentclass[11pt,floats,prd,aps,twocolumn,amssymb,nofootinbib,letterpaper, superscriptaddress,showkeys]{revtex4-2}
\usepackage{amssymb}
\usepackage{graphics}
\usepackage{amsmath}
\usepackage{amsfonts}
\usepackage{bm}
\usepackage[]{latexsym}
\usepackage{cellspace}
\usepackage{mathtools}
\usepackage{amstext}
\usepackage{stackrel}
\usepackage{graphics}
\usepackage{graphicx}
\usepackage[utf8]{inputenc}
\usepackage{blindtext}
\usepackage{float}
\restylefloat{table}
\usepackage{booktabs}
\usepackage{enumitem}
\usepackage{amssymb,amsbsy,epsfig,graphicx}
\usepackage{color}
\usepackage{etoolbox} 
\usepackage{lipsum} 
\usepackage{eufrak}
\usepackage{verbatim}
\usepackage{bm}
\usepackage{hyperref}
\usepackage{cleveref}
\usepackage{natbib}
\usepackage{xcolor}
\hypersetup{pdfstartview=FitV,colorlinks=true,linkcolor=midblue,citecolor=midblue,filecolor=midblue,urlcolor=midblue}

\definecolor{midblue}{rgb}{0,0,0.5}
\usepackage[left=1.5cm,right=1.5cm,top=3.2cm,bottom=3.2cm]{geometry}

\def\actaa{\ref@jnl{Acta Astron.}}      

\newcommand{\be}{\begin{equation}}\newcommand{\ee}{\end{equation}}
\newcommand{\bea}{\begin{eqnarray}}\newcommand{\eea}{\end{eqnarray}}
\newcommand{\brr}{\begin{array}}\newcommand{\err}{\end{array}}
\newcommand{\bit}{\begin{itemize}}\newcommand{\eit}{\end{itemize}}
\newcommand{\ben}{\begin{enumerate}}\newcommand{\een}{\end{enumerate}}

\newcommand{\ba}{\begin{array}}
	\newcommand{\ea}{\end{array}}

\begin{document}
	
	\title{Spontaneous Lorentz symmetry breaking effects on GRBs jets arising from neutrino pair annihilation process near a  black hole}
	

	
	\author{\textbf{Mohsen Khodadi}}\email{m.khodadi@hafez.shirazu.ac.ir}
	\affiliation{Department of Physics, College of Sciences, Shiraz University, Shiraz 71454, Iran}
	\affiliation{Biruni Observatory, College of Sciences, Shiraz University, Shiraz 71454, Iran}
	
	\author{\textbf{Gaetano Lambiase}}\email{lambiase@sa.infn.it}
	\affiliation{Dipartimento di Fisica ``E.R Caianiello'', Università degli Studi di Salerno, Via Giovanni Paolo II, 132 - 84084 Fisciano (SA), Italy}
	\affiliation{Istituto Nazionale di Fisica Nucleare - Gruppo Collegato di Salerno - Sezione di Napoli, Via Giovanni Paolo II, 132 - 84084 Fisciano (SA), Italy}
	
	\author{\textbf{Leonardo Mastrototaro}}\email{lmastrototaro@unisa.it}
	\affiliation{Dipartimento di Fisica ``E.R Caianiello'', Università degli Studi di Salerno, Via Giovanni Paolo II, 132 - 84084 Fisciano (SA), Italy}
	\affiliation{Istituto Nazionale di Fisica Nucleare - Gruppo Collegato di Salerno - Sezione di Napoli, Via Giovanni Paolo II, 132 - 84084 Fisciano (SA), Italy}
	
	\date{\today}
	\def\be{\begin{equation}}
		\def\ee{\end{equation}}
	\def\al{\alpha}
	\def\bea{\begin{eqnarray}}
		\def\eea{\end{eqnarray}}

	\begin{abstract}
		The study of neutrino pair annihilation into electron-positron pairs ($\nu{\bar \nu}\to e^-e^+$) is astrophysically well-motivated because it is a possible powering mechanism for the gamma-ray bursts (GRBs). In this paper, we estimate the gamma-ray energy deposition rate (EDR) arising from the annihilation of the neutrino pairs in the equatorial plane of a slowly rotating black hole geometry modified by the broken Lorentz symmetry (induced by a background bumblebee vector field). More specifically, owing to the presence of a dimensionless Lorentz symmetry breaking (LSB) parameter $l$ arising from nonminimal coupling between the bumblebee field with nonzero vacuum expectation value and gravity, the metric solution in question differs from the standard slowly rotating Kerr black hole.
		By idealizing the thin accretion disk temperature profile in the two forms of isothermal and gradient around the bumblebee gravity-based slow rotating black hole, we investigate the influence of spontaneous LSB on the $\nu{\bar \nu}$-annihilation efficiency.
		For both profiles, we find that positive values of LSB parameter $l>0$ induce an enhancement of the EDR associated with the neutrino-antineutrino annihilation. Therefore, the process of powering the GRBs jets around bumblebee gravity modified slowly rotating geometry is more efficient in comparison with standard metric. Using the observed gamma-ray luminosity associated with different GRBs types (short, long, and ultra-long), we find, through the analysis of the EDR in the parameter space $l-a$ ($a^2\ll1$), some allowed ranges for the LSB parameter $l$.
		
	\end{abstract}
	
	\vskip -1.0 truecm
	\maketitle
	
	\section{Introduction}
	\label{Introduction}
	The Lorentz invariant (LI), a continuous symmetry representing the physical results independent of the boost and rotation, is one of the key characteristics at the heart of modern physics. Conventionally, LI is related to the scale-free nature of spacetime. Experimentally this is well-supported, and there is no reason to believe that LI is broken at the currently accessible energies \cite{Mattingly:2005re, Liberati:2013xla} (for a detailed list of results see \cite{Kostelecky:2008ts}). However, it is expected that the coupling constants in quantum field theories are energy-dependent. Namely, unlike low energy scales, the coupling constants addressing the Lorentz-breaking terms become significant at high energy. Searching for a Lorentz symmetry breaking (LSB) is one of the hot topics in theoretical physics since it is strictly related to fundamental issues, such as quantum gravity and string theory \cite{Kostelecky:1990pe, Kostelecky:1994rn}. Even though the nature of quantum gravity is not determined, it is well-known that space and time should break down at small distances (around Planck's length), where they can no longer be treated as a classical continuum \cite{Amelino-Camelia:2008aez}.  Therefore, there is a pervasive belief that LI cannot be a well-established symmetry at all scales of nature, and it becomes invalid by approaching the fundamental scales \cite{Collins:2006bw}. In the modern analyses for detecting experimental deviations from LI, a phenomenological framework, known as Standard-Model Extension (SME), is utilized in which LI violating effects are introduced by spontaneous symmetry breaking \cite{Colladay:1998fq}. SME is an effective field theory for the development of low-energy experiments containing all possible LI and Charge-Parity-Time (CPT) violating coefficients\footnote{For the deep connection of LI with CPT theorem, in the case of violating the former, the latter is challenged, too.} that include not only Special Relativity but the Standard Model and General Relativity as well \cite{Colladay:1996iz}.  Overall, the presence of both Nambu-Goldstone and massive Higgs bosons in theories with spontaneous LSB provide us with rich scenarios for exhibiting multiple phenomenological effects \cite{Bluhm:2004ep, Kostelecky:2005ic}. In this respect, by implementing the spontaneous LSB into a curved spacetime via background vector fields, it became possible to construct some well-known alternatives to General Relativity such as the Einstein-Aether theory \cite{Jacobson:2000xp} and the Bumblebee Gravity (BG) model\footnote{An advantage of BG model, compared to Einstein-Aether theory, is that it has no difficulties of Einstein-Aether in the perturbative description \cite{Petrov:2020wgy}.} \cite{Bluhm:2004ep,Bertolami:2005bh}.
	
	Concerning the BG, to which we are interested in, the LSB mechanism occurs owing to the nonzero vacuum condensation of a vector field (the bumblebee field), indicating a preferred frame \cite{Kostelecky:2003fs, Bluhm:2007xzd}. The most peculiar feature of the BG model that has attracted a lot of attention is that it has no local $U(1)$ gauge symmetry, but the propagation of massless vector modes is still allowed \cite{Bluhm:2008yt}. Indeed, in the BG model, the massless photon modes arise as Nambu-Goldstone bosons when the spontaneously LSB occurs. Bumblebee vector fields can be a source of cosmological anisotropies since they generate a preferred axis \cite{Liang:2022hxd}. This means, in turn, that a fraction of the anisotropies observed in the universe can be ascribed to LI violation \cite{Maluf:2021lwh}. In this regard, it has been recently derived, by implementing the big bang nucleosynthesis and gravitational baryogenesis governing in early universe within the BG-based cosmology, some tight constraints for the Lorentz-violating bumblebee vector field \cite{Khodadi:2022mzt}.
	Ref. \cite{Capelo:2015ipa}  shows that spontaneous LSB arising from the bumblebee field in the background can describe the dark energy issue. Unlike general SME that suffers from the lack of an exact gravitational solution, in the BG model it is possible to find a Schwarzschild-like solution \cite{Casana:2017jkc} (see also the spherically symmetric solution in recent Ref. \cite{Filho:2022yrk} which shows the effect of the LSB on both the temporal and radial components), as well as a Kerr-like black hole \cite{Ding:2019mal} (traversable wormhole solutions have been found in \cite{Ovgun:2018xys}). 
A Kerr-like black hole solution derived in \cite{Ding:2019mal} was re-evaluated in \cite{Maluf:2022knd}. The conclusion is that, at the present, there is no yet full rotating black hole solution for the Einstein-bumblebee theory. The case of slowly rotating metric has been derived in \cite{Ding:2020kfr}, whose validity is confirmed in \cite{Maluf:2022knd}, too.
	
	In these solutions, the LSB arises from a nonzero vacuum expectation value (VEV) of the bumblebee vector field coupled to the spacetime curvature. These solutions allow to study various aspects of the role of spontaneously LSB on the physics of compact objects, such as black holes and wormholes (e.g., see Refs. \cite{Gomes:2018oyd,Oliveira:2018oha, Liu:2019mls, Maluf:2020kgf, KumarJha:2020ivj, Jha:2020pvk, Khodadi:2021owg, Jha:2021eww, Jiang:2021whw, Wang:2021gtd, Khodadi:2022dff, Gu:2022grg}).  It would be interesting to mention that the study of causality in Lorentz-violating models is a relevant topic.  In Refs. \cite{Santos:2014nxm, Jesus:2020lsv}, the authors have studied the Gödel-type solution for the BG model.	In the light of confronting this model with astrophysical bodies, such as stars, were derived tight constraints for the LSB parameter \cite{Paramos:2014mda}.
	
	The aim of this paper is to study the modification induced by BG on the energy deposition rate (EDR) of neutrinos emitted from a thin accretion disk~\cite{Asano:2000dq,Asano:2000ib}. More exactly, if the disk is hot enough and the accretion rate of black hole obeys the condition $\dot{M}\sim (0.1-1) M_{\odot}$ s$^{-1}$, then 
	the disk is a source of neutrinos ($\nu$) and anti-neutrinos ($\bar{\nu}$), which partially annihilate above the disk and convert into electron ($e^{-}$) and positron ($e^{+}$), as follows \cite{Popham:1998ab, Birkl:2006mu, Chen:2006rra, Zalamea:2010ax}
	\begin{equation}\label{nunuee}
		\nu{\bar \nu}\to e^+ e^- \,.
	\end{equation}
	This process has significant consequences for the cosmological gamma-ray bursts (GRBs) jets, which are the most luminous objects in the universe. Accretion disks around black holes are the favourite candidates for the central engine of GRBs. Such a configurations are formed by the merging of compact objects (neutron-neutron stars, black hole-neutron stars) or by supernova.
	The hot accretion disk in these systems is the source of neutrinos/antineutrinos, and their annihilation into electrons/positrons may power GRBs. 
	However, in order that the relativistic fireball produces GRBs, the fireball must contain an extremely small amount of the baryon density. The latter, above the accretion disk, is low near the rotation axis so that the neutrino pair annihilation has the possibility of producing a clean fireball which allows to solve the baryon contamination problem, which hinders the creation of relativistic shocks and the emission of gamma-rays (see \cite{Asano:2000dq} and references therein).  More exactly, the jets produced via neutrino annihilation, in essence, are cones relatively free of baryons. Note, finally, that in the processes of emission of neutrinos from the hot accretion disk and pair annihilation are important the relativistic effects that essentially take into account the gravitational redshift, the bending of the neutrino trajectories, and the redshift due to rotation. The EDR is affected by these effects when Kerr geometry is modified by LI violating corrections, as will see in what follows.

	GRBs, in essence, are powerful cosmic explosions of the characteristic duration of seconds. Commonly are labeled into two classes: short and long GRBs for timescales, $\sim 1$ s and $(1-100)$ s, respectively \cite{Woosley:1993wj,Galama:1998ea,Stanek:2003tw}. Despite the uncertainty about the origin of these two categories, the evidence suggests that the former most likely comes from the merger of compact binaries such as a double neutron star and/or a binary system, including a neutron star and black hole \cite{Eichler:1989ve,Narayan:1992iy,Nakar:2007yr}, while the collapse of the core of massive stars (Wolf–Rayet) can be the origin of the latter \cite{Woosley:1993wj}. Based on the observed gamma-ray luminosity $L$, the luminosity of the long-duration type of GRBs should not exceed $L\sim 10^{53}$ erg/s (more accurate $10^{52-53}$ erg/s) \cite{Bloom:2003eq}. Overall, the order of magnitude of the luminosity of short and long GRBs is expected to be the same \cite{Leng:2014dfa}. 
	%
	In recent years, however, a new population of ultra-long GRBs with timescale duration $\sim 10^{3-4}$ s that reduces the luminosity up to $\sim 10^{49-50}$ erg/s
	(e.g, see Refs. \cite{Thone:2011yf,Gendre:2012wj,Levan:2013gcz}) has been investigated. 
	The relation of the maximum energy of a neutrino-powered jet as a function of the burst duration shows that the energy deposition falls down rapidly as the burst lasts longer \cite{Leng:2014dfa}. 
	
	Observations indicate that the process (\ref{nunuee}), due to the creation of relativistic $e^{\mp}$-dominated jets, can be a possible candidate to explain GRBs observed from galaxies containing the supermassive black hole in their center, e.g., see \cite{Zalamea:2010ax}. In \cite{Co86,Co87,Goodman:1986we,Eichler:1989ve,1993AcA....43..183J} it has been shown that the process (\ref{nunuee}) can deposit an energy $\gtrsim 10^{51}$ erg above the neutrino-sphere of a type II supernova \cite{Goodman:1986we}. In \cite{Salmonson:1999es, Salmonson:2001tz} it has been shown that taking into account the effects of the strong gravitational field in a Schwarzschild spacetime, the efficiency of the process (\ref{nunuee}), for collapsing neutron stars, enhances (up to a factor of $30$) compared to the Newtonian case. 
	The same analysis around a thin and isothermal accretion disk for a Schwarzschild or Kerr metric was performed in \cite{Asano:2000ib,Asano:2000dq},
	%
	%
	The neutrino annihilation luminosity from the disk has been also calculated, e.g., see \cite{Mallick:2008iv,Mallick:2009nvq,Chan:2009mw,Kovacs:2009dv,Kovacs:2010zp,Zalamea:2008dq,Harikae:2010yt}.
	Time-dependent models of black hole accretion disks, such as remnants of neutron-star mergers or collapse engines, have been investigated, for example, in Refs. \cite{Harikae:2010yt,Ruffert:1998qg,Popham:1998ab,DiMatteo:2002iex,Fujibayashi:2017xsz,Just:2015dba,Foucart:2018gis,Foucart:2020qjb}. 
	The principal output of these studies is that the neutrino-pair annihilation process, when analyzed in curved background described by General Relativity, is not efficient enough to power GRBs. There is another scenario for energy extraction from disk or black hole to launch the GRBs jets, the well-known magnetohydrodynamical (e.g., \cite{Katz:1997bh,Meszaros:1996ww}). According to this scenario, the Blandford-Znajek process is a more promising mechanism for launching jets. However, the main issue of this energy extraction model is whether (yet has been not proven) the magnetic flux arising from the collapse of the star is sufficient to power the jet or not \cite{Komissarov:2009dn}. Note that throughout this paper, we will only address the neutrino effects without considering the contribution of other potential forms of energy deposition, such as Blandford-Znajek and magnetic reconnection.
	
	The process (\ref{nunuee}), with neutrino-antineutrino emitted from the surface of a neutron star, has been also investigated in the framework of extended theories of gravity  \cite{Lambiase:2020iul, Lambiase:2020pkc, Lambiase:2022ywp}.
	By admitting this idea that the environment around a black hole potentially is a cleaner place for the launch of a relativistic jet,  we consider the BG-based rotating black hole with broken LI surrounded by a thin accretion disk from which neutrinos are emitted \cite{Asano:2000dq}. Inspired by 
	Refs. \cite{Asano:2000dq, Asano:2000ib,10.1143/PTPS.136.235}, we assume an idealized, semi-analytical, stationary state model, independent of details regarding the disk formation. Note that the self-gravitational effects are not taken into account, and the disk is described by an inner and outer edge.
	
	The plan of our work is as follows. In Sec.~\ref{BMOD} we overview the modified slowly rotating Kerr black hole solution addressed by the BG model.
	In Sec.~\ref{Formulation} we present the model used for computing the energy deposition from the thin disk. In Sec.~\ref{Results} we characterize the effects of the theories beyond General Relativity on the EDR of neutrino pair annihilation near the rotational axis of the gravitational source. We shall consider two profiles for the disk temperature: $T=const$ and $T\propto r^{-1}$.
	Finally, in Sec.~\ref{Conclusion} we summarize our results.
	
	\begin{table*}[t]
		\begin{tabular}{|c|c|c|}
			\hline 
			Allowed range of  $l$ and $a$ & Scenario& Refs. \\
			\hline
			$(-1,0.6]$ for $a \in [0.5,1)$&  BG with with Kerr-Sen-like solution in light of Event Horizon Telescope ($M87^*$) \footnote{Note to this point is essential that the shadow of BG with Kerr-like black hole solution introduced in \cite{Ding:2019mal} is not distinguishable from its standard counterpart, see \cite{Vagnozzi:2022moj} for more details.} & \cite{Jha:2021eww} \\  \hline
			$[-0.23,0.06]$ for $a \in [0.28,0.31]$&  BG with Kerr-like solution in light of quasi-periodic oscillations & \cite{Wang:2021gtd} \\
			\hline
			$[-0.56,6.5]$ for $a \in [0.32,0.81]$&  - & \cite{Wang:2021gtd} \\
			\hline
			$[-0.7,10.8]$ for $a \in [0,4]$&  - & \cite{Wang:2021gtd} \\
			\hline
		\end{tabular}
		\caption{ The allowed ranges of $l$, that have been obtained via confronting the BG black hole with observational data. The constraint reported in the third to fifth rows, respectively come from the three observational data: GRO J1655-40 \cite{Motta:2013wga}, XTE J1550-564 \cite{Orosz:2011ki}, and GRS 1915+105 \cite{Reid:2014ywa}, within $1\sigma$ level.}
		\label{tab:rangel}
	\end{table*}
	
	\section{modified slowly rotating Kerr black hole solution with a background bumblebee field}
	\label{BMOD}
	
	In this Section, we shortly review the slowly rotating Kerr black hole solution obtained from the nonminimal coupling of the background bumblebee field to gravity. For that, the spontaneous LSB occurs in a curved spacetime and the metric tensor must couple to the vector field. This leads to the bumblebee action \cite{Bluhm:2004ep,Bertolami:2005bh} (in units $c=G_N=1$)
	\begin{eqnarray}\label{BAction}
		S &=&\int d^{4}x\sqrt{-g}\bigg( \frac{1}{16\pi}\left(R+\xi B^{\mu
		}B^{\upsilon}R_{\mu\nu}\right) \\
		& & \qquad
		-\frac{1}{4}B^{\mu\nu}B_{\mu\nu}-V\left(
		B^{\mu}\right)  \bigg)\,.
		\nonumber
	\end{eqnarray}
	Here $B_\mu$ is the bumblebee vector field with mass dimension $M$,
	$B_{\mu\nu}=\partial_{\mu}B_{\nu}-\partial_{\nu}B_{\mu}$ the bumblebee-field strength with mass dimension $M^{2}$, $\xi$ the nonminimally coupling constant between
	the background bumblebee field and gravity with mass dimension $M^{-2}$, and finally,  $V\left(B^{\mu}\right)$ the potential defined as
	\[
	V\left(B^{\mu}\right)=B_{\mu}B^{\mu}\pm b^{2}\,.
	\]
	The potential $V(B^{\mu})$ is such that $B_\mu$ may acquire a nonzero VEV $\langle B^{\mu}\rangle=b^{\mu}$ (spontaneous LSB in the gravitational sector \cite{Kostelecky:2003fs, Bluhm:2004ep}).
	
	The slowly rotating black hole metric derived from the BG in the Boyer-Lindquist coordinates $x^\mu=(t,r,\theta,\phi)$ is given by (see Ref. \cite{Ding:2020kfr} for details)
		\begin{eqnarray}
			ds^{2}&=&-\left(  1-\frac{2M}{r}\right)  dt^{2}-\frac{4Ma\sin^{2} \theta }{r}dtd\phi \nonumber\\ 
			&&+\frac{r^{2}}{\tilde{\Delta}}dr^{2}
			+r^{2}d\theta^{2}+r^2\sin^{2}  \theta d\varphi^{2}, \label{metricBmod}
		\end{eqnarray}
		where
		\begin{eqnarray}
			\tilde{\Delta} &=& \frac{r^{2}-2Mr}{l+1}\,,~~~l\neq-1 \label{Deltametr}   
	\end{eqnarray}
	It gets modified by a soft deviation from the standard slowly rotating Kerr due to the appearance of dimensionless LSB parameter ($l=\xi b^2$), so that the final form of the metric tensor reads
		\begin{equation}\label{metricg}
			g_{\mu\upsilon}=\left(
			\begin{array}[c]{cccc}
				-\left(  1-\frac{2M}{r}\right)  & 0 & 0 & -\frac{2Ma \sin^{2}\theta}{r} \\
				0 & \frac{r^{2}}{\tilde{\Delta}} & 0 & 0  \\
				0 & 0 & r^{2} & 0  \\
				-\frac{2Ma\sin^{2}\theta}{r} & 0 & 0 & r^2\sin^{2} \theta
			\end{array}\right)\,,
		\end{equation}
		As it is clear, the parameter $l$ leaves a distinguishable imprint on the metric via $\tilde{\Delta}$. As a result, the underlying metric differs from the standard slowly rotating Kerr metric. In other words, the BG-based metric at hand differs softly from the standard slowly rotating Kerr metric by a factor $(l+1)$ in the standard definition of component of $g_{11}$ i.e., $\frac{l+1}{1-2M/r}$.

	Due to the play of the constructive role of LSB parameter $l$ on the results of our analysis, it is worthwhile to discuss it in a bit of detail. In essence, the sign of $l$ comes from the sign of the nonminimal coupling constant ($\xi$) between the background bumblebee vector field and gravity. Since there is no consensus in the literature, we deal with both signs negative \cite{Wang:2021gtd} and positive \cite{Paramos:2014mda, Casana:2017jkc}. Given the fact that the BG is a subclass of SME, it is shown that via Parameterized Post-Newtonian (PPN) analysis, the spacelike background bumblebee vector field $b_{\mu}$ is matched to dimensionless tensor $s^{\mu\nu}$ in SME, see Ref. \cite{Bailey:2006fd} for more details. Namely, the LSB parameter $l$ can be limited, via constraints imposed on Lorentz violating coefficients of the SME, to find most of the upper bounds extracted on the different combinations of $s^{\mu\nu}$  (see the review paper \cite{Hees:2016lyw}). In the recent paper \cite{Khodadi:2022pqh} there is a summarized list of the most important physical frameworks to derive upper bounds on the $s^{\mu\nu}$. Moreover, stringent constraints on $l$ have been directly inferred in Refs. \cite{Paramos:2014mda, Casana:2017jkc} by considering BG in the framework of astrophysics and some classic tests. In this regard, it is recommended to visit some more recent works such as \cite{Maluf:2021lwh,Khodadi:2022mzt} as well.
	The common feature of all these constraints, whether directly or through connection with Lorentz violating coefficients of the SME, is that they have been derived in the weak-field regimes with the gravitational redshift $\ll 1$. However, the Lorentz violation effects appear at fundamental scales, as pointed out in the Introduction. The environment around compact objects, such as a black hole, no longer belongs to the weak-field regime since its redshift is $\gtrsim 1$.  So it is reasonable to imagine that by increasing the redshift of the gravitational field under investigation (as around black holes), the current constraints derived in the weak-gravity regime may change \cite{Psaltis:2008bb}. In light of these points, one can safely relax the above-mentioned constraints on $l$ for the framework at hand. This also occurs in the frameworks reported in Table.~\ref{tab:rangel} where some BG black hole scenarios are directly compared with observational data. 
	%
	%
	It might be interesting to stress that in Ref. \cite{Gu:2022grg} were used newly the blurred reflection traits in the X-ray spectra of galactic black hole EXO 1846–031 to constrain $l$. Despite the lack of success to do it due to the degeneracy between the rotation parameter of the black hole and the LSB parameter, it is expected to fix this problem by combining other observations in future analysis. An important point to note about the above-mentioned constraints for $l$ in the framework of BG is that they come from taking the Kerr-like solution \cite{Ding:2019mal}, which is just valid in the slow rotating limit ($a^2\ll1$) \cite{Maluf:2022knd,Ding:2020kfr}. It means that these constraints can not be reliable beyond slow-rotating approximation.

	\section{Energy deposition rate from $\nu{\bar \nu}$ annihilation}
	\label{Formulation}
	
	Let us consider a black hole with a thin accretion disk around it that emits neutrinos \cite{Asano:2000dq}. We will confine ourselves to the case of an idealized, semi-analytical, stationary state model, which is independent of details regarding the disk formation. The disk is described by an inner and outer edge, with corresponding radii defined by $R_{\mathrm{in}}$ and $R_{\mathrm{out}}$, respectively. Self-gravitational effects are neglected. We consider the generic metric
	\begin{equation}
		g_{\mu\nu}= \left(\begin{matrix} g_{00}& 0& 0& &g_{03}\\0& g_{11}& 0& &0\\0& 0& g_{22}& &0\\g_{03}& 0& 0& &g_{33}
		\end{matrix}\right) \,.
	\end{equation}
	
	The Hamiltonian of a test particle reads
	\begin{equation}
		2\mathcal{H}=-E\dot{t}+L\dot{\phi}+g_{11}\dot{r}^2=\delta_1 \,,
	\end{equation}
	where $\delta_1=0, 1$ refers to null geodesics and massive particles, respectively, $E$ and $L$ are the energy and angular momentum of the test particles moving around the rotational axis of the black hole. The non-vanishing components of the 4-velocity are \citep{Prasanna:2001ie}
	\begin{align}
		U^{3}&=\dot{\phi}=E\left(\frac{L}{E}+\frac{1}{2}\frac{g_{03}}{g_{00}}\right)\left(g_{33}-\frac{1}{2}\frac{g_{03}^2}{g_{00}}\right)^{-1} \,, \nonumber \\
		U^0&=\dot{t}=-\frac{E}{g_{00}}
		\left[1+\frac{g_{03}}{2}
		\left(\frac{L}{E}+
		\frac{g_{03}}{2g_{00}}\right)
		\left(g_{33}-\frac{g_{03}^2}{2g_{00}}\right)^{-1}\right] \nonumber \\
		\dot{r}^2&=
		\frac{E\dot{t}-L\dot{\phi}}
		{g_{11}} \,. \nonumber
	\end{align}
	We are interested in the energy deposition rate near the rotational axis at $\theta=0^{\circ}$. We use the value $\theta=0^{\circ}$ for evaluating the energy emitted in a half cone of $\Delta \theta\sim10^{\circ}$. The accretion disk extends from $R_{in}=2R_{\mathrm{ph}}$ to $R_{out}=30M$, with $R_{\mathrm{ph}}$ the photosphere radius. Moreover, it can be shown that the following relation for the impact parameter holds~\cite{Asano:2000dq}
	\begin{equation}
		\rho_\nu=\sqrt{g_{00}(r_0,0)g_{22}(r_0,0)} \,,
		\label{rho}
	\end{equation}
	with $r_0$ the nearest position between the particle and the centre before arriving at $\theta=0$. Finally, from the metric  (\ref{metricg}), the equation of the  trajectory  becomes \cite{Asano:2000dq}
	\begin{equation}
		\int\frac{d\theta}{\sqrt{1-(\tilde{a}/\rho_\nu)^2\sin^2\theta}}=
		\int \frac{dr'}{\sqrt{\frac{g_{22}^2(r',0)}{\rho^2_{\nu}}-\frac{g_{22}(r',0)}{g_{11}(r',0)}}}\,. \nonumber
	\end{equation}
	In this relation one takes into account that the neutrinos are emitted from the position $(R,\pi/2)$, with $R\in [R_{in},R_{out}]$, and arrive at $(r,0)$. The energy deposition rate of neutrino pair annihilation is given by \cite{Asano:2000dq}
	\begin{equation}
		\frac{dE_0(r)}{dtdV}=\frac{21\pi^4}{4}\zeta(5)KG_F^2T^9_{\mathrm{eff}}(R_{2R_{ph}})F(r, T_0) \,,
		\label{trajectory}
	\end{equation}
	where $G_F$ is the Fermi constant, $k$ is the Boltzmann constant, $T_{\mathrm{eff}}(2R_{\mathrm{ph}})$ is the effective temperature at radius $2R_{ph}$ (the temperature observed in the comoving frame),
	\begin{equation}
		K=\frac{1\pm 4\sin^2\omega_W+8\sin^4\omega_W}{6\pi} \,,
	\end{equation}
	with the $+$ sign for $\nu_e$ and the $-$ sign for $\nu_{\mu/\tau}$, $F(r, T_0)$ is reported in the Appendix (Eq. \ref{F(r)}),
	%
	%
	%
	with $T_0$ the temperature observed at infinity
	\begin{align}
		&T_0(R)=\frac{T_{\rm{eff}}\left(R,\frac{\pi}{2}\right)}{\gamma}\sqrt{g_{00}\left(R,\frac{\pi}{2}\right)-\frac{g_{03}^2(R,\frac{\pi}{2})}{g_{33}(R,\frac{\pi}{2})}} \,\ , \\
		&\gamma=\frac{1}{\sqrt{1-v^2/c^2}} \,\ , \\
		&\frac{v^2}{c^2}=\frac{g_{33}^2(r,\pi/2)\left(\Omega_K-\omega\right)^2}{g_{03}^2(r,\pi/1)-g_{00}(r,\pi/2)g_{33}(r,\pi/2)} \,\ , \\
		&\Omega_K=\frac{-g_{03,r}+\sqrt{(g_{03,r})^2-g_{00,r}g_{33,r}}}{g_{33,r}}\Big|_{(r,\pi/2)} \,\ , \\
		&\omega=-\frac{g_{03}(r,\pi/2)}{g_{33}(r,\pi/2)} \,\ .
	\end{align}
	where $T_{\mathrm{eff}}$ is the effective temperature measured by a local observer and all the quantities are evaluated at $\theta=\pi/2$. In the treatment we will ignore the reabsorption of the deposited energy by the black hole and we will consider the case of isothermal disk, that is
	\[
	T_{\mathrm{eff}}=const\,,
	\]
	and the case of a gradient temperature \cite{Asano:2000dq}, for which $T_{\mathrm{eff}}$, in the simplest and acceptable model, is given by (for details, see \cite{10.1143/PTPS.136.235})
	\begin{equation}\label{tdippr}
		T_{\mathrm{eff}}(r)= \frac{2R_{ph}}{r} \,.
	\end{equation}

	\section{Applications to the Bumblebee metric}
	\label{Results}
	
	In this section, we calculate the emitted energy with the procedure shown in Sec.~\ref{Formulation} for the bumblebee metric in Eq.~(\ref{metricg}). We analyze two different cases, corresponding to the isothermal model and the temperature gradient model. We find an enhancement of the EDR for positive values of the LSB parameter $l$ ($l>0$) entering Eq. (\ref{metricg}).
	
	\subsection{Isothermal model}
	
	For our analysis, it turns out convenient to define the function
	
	\begin{equation}\label{G(r)Gae}
		G(r)=F(r, T_0)\frac{r^2+a^2}{4M^2} \,,
	\end{equation}
	where $F(r, T_0)$ is given in Eq. (\ref{F(r)}).
	The function $G(r)$ is essential for evaluating the EDR and, therefore, the energy viable for a GRB explosion.
	The EDR is estimated for the infinitesimal angle $d\theta$ taking into consideration a characteristic angle $\Delta \theta \simeq 10^{\circ}$ and temperature $T_{\rm{eff}}= 10 \mathrm{MeV}$. The explicit EDR expression is given by \cite{Asano:2000dq}
	
	\begin{widetext}
		\begin{equation}
			\frac{dE_0}{dt}\simeq 4.41\times 10^{48}\left(\frac{\Delta \theta}{10^\circ}\right)^2\left(\frac{T_\mathrm{eff}(R_{\mathrm{in}})}{10~\mathrm{MeV}}\right)^9\left(\frac{2M}{10~\mathrm{km}}\right)\int_{R_{\mathrm{in}}}^{R_{\mathrm{out}}}\frac{G(r)}{2M}dr~\mathrm{erg~s^{-1}} \,\ .
			\label{value}
		\end{equation}
		
	\end{widetext}

	\begin{figure}[t]
		\centering
		\includegraphics[scale=0.7]{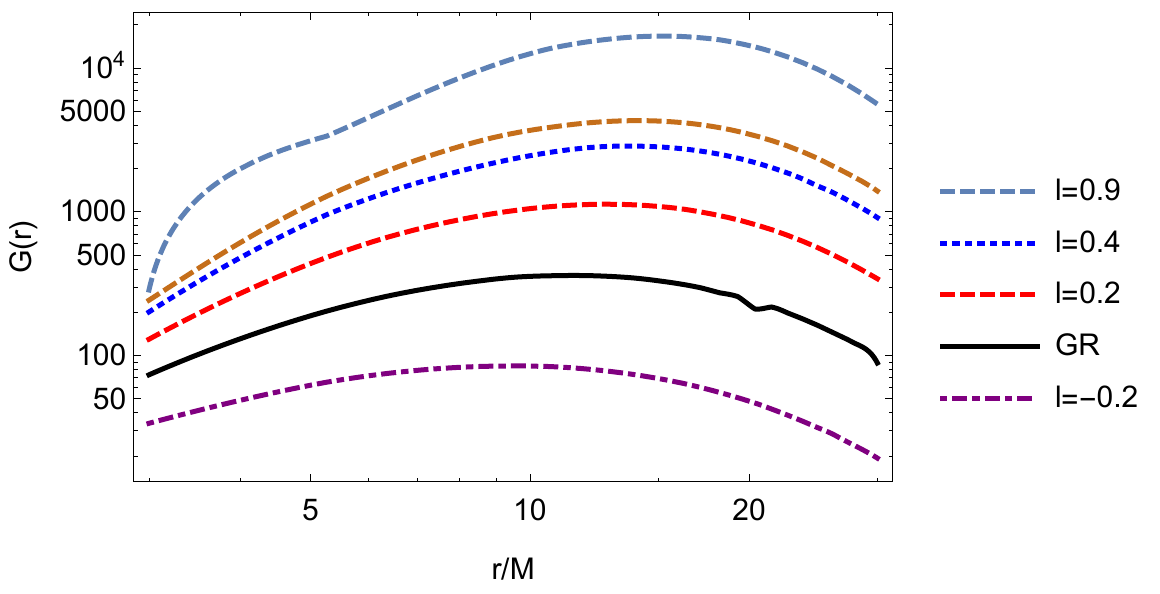}
		\includegraphics[scale=0.7]{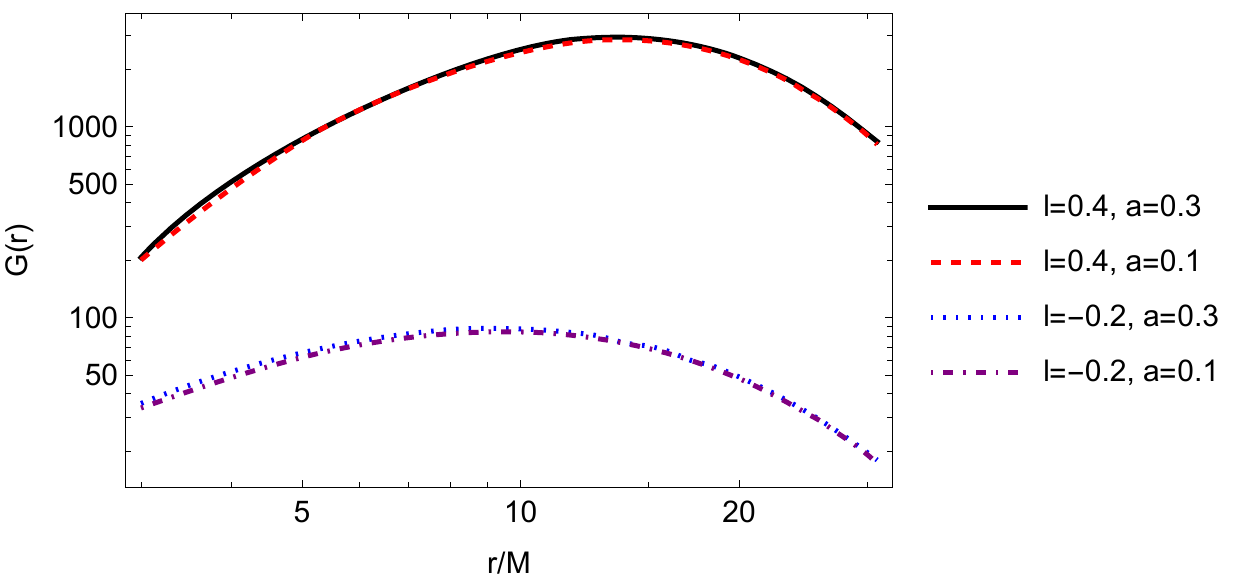}
		\caption{Plot of $G(r)$ against $r/M$ for the isothermal disk. \textbf{Upper graph:} Different values of $l$ with fixed value $0.3$ for the slowly rotating parameter $a$.  \textbf{Lower graph:} Different values of the spin parameter $a$ with $l<0$ and $l>0$.}
		\label{Gr}
	\end{figure}
	Before going into the analysis of EDR, a comment is in order.
	The function $F(r,T_0)$ defined in Eq. (\ref{F(r)}), is proportional to $g_{11}^4$. From (\ref{metricg}) it follows that $g_{11}\propto 1/\tilde{\Delta}$ and $\tilde{\Delta}\propto 1/(1+l)$, resulting in $g_{11}\propto (1+l)$. Therefore, the enhanced factor $(1+l)^4$ appears in the computation of the EDR. Moreover, the LSB parameter $l$ also contributes, via the modifications in the metric (\ref{metricg}), to the trajectory equation leading to important changes in the neutrino angular momentum at the rotational axis.
	
\begin{figure}[t]
		\centering
		\includegraphics[scale=0.8]{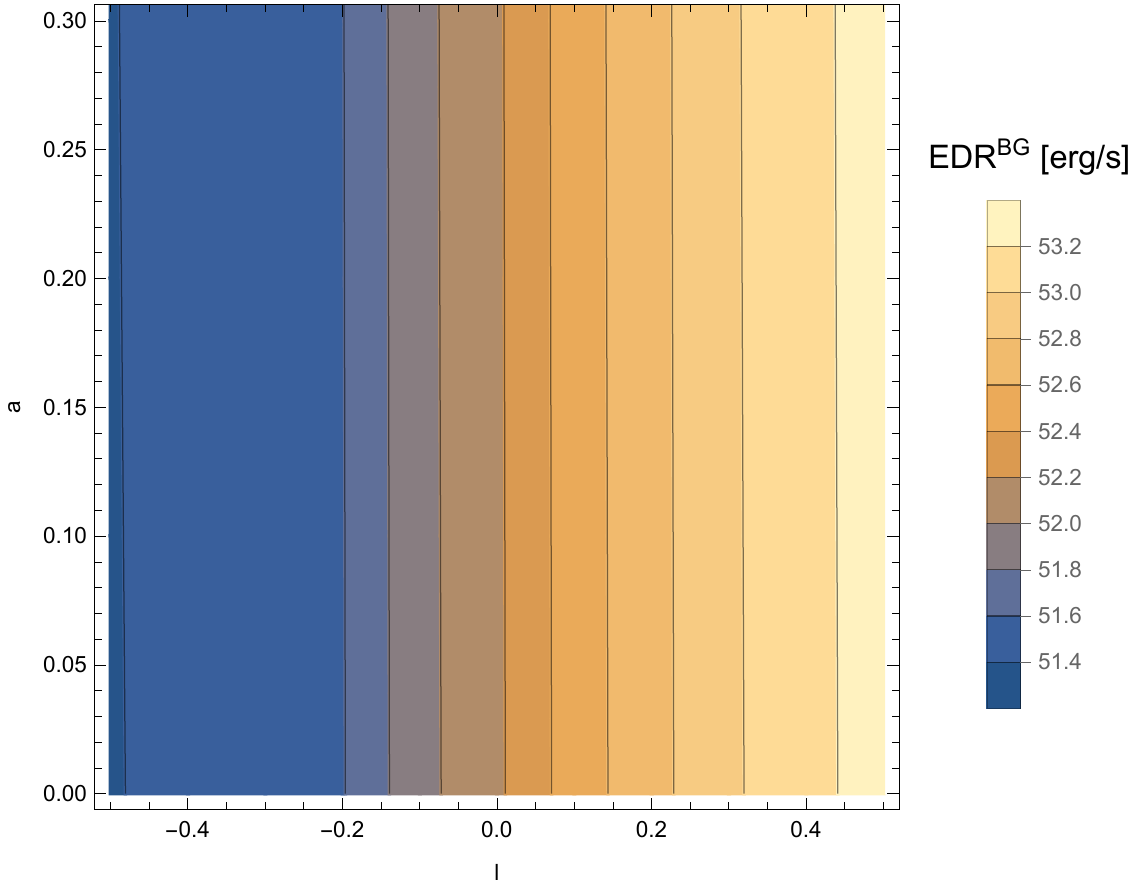}
		\caption{A contour plot of the parameter space $l-a$ showing the EDR$^{BG}$ related to BG-based slowly rotating Kerr black hole in isothermal disk model. Here the color scale written in the logarithmic form $\log 10^{n}$ (the range of $n$ is from values lesser than $52$ to beyond $53$ ).}
		\label{Con1}
	\end{figure}

	In Fig.~\ref{Gr} we show the behavior of $G(r)$ in BG metric and its standard counterpart, as well. In the upper graph, we plot curves with $l<0$, $l>0$ and $l=0$ by fixing the spin parameter to $a=0.3$. As a common feature in both cases,  $G(r)$ initially increases with the distance, reaching a maximum value, and then decreases with the distance due to the interplay between temperature and red-shift effects. However, we see that the presence of the LSB parameter, with positive values, increases $G(r)$ compared to $l=0$, while for $l<0$ the function $G(r)$ decreases. By drawing the lower graph in Fig.~\ref{Gr}, we are interested in investigating the role of the slowly rotating parameter $a$ on the behavior of $G(r)$ (subsequently of the EDR) in the interplay with $l<0$ and $l>0$. As one can see, the spin parameter $a$ does not affect the curves with both negative and positive values of the LSB parameter $l$.
		This means that for the isothermal disk model in an LSB-based metric such as BG, the rotation of the black hole has no effective role in the EDR. 
		This can be seen more clearly in the contour plot of the parameter space $l-a$ in Fig.~\ref{Con1}. As it is evident, the EDR$^{BG}$ increases by moving from $l<0$ to $l>0$, independently of the value of the spin parameter. Indeed, the main contribution to the increase of the EDR, which makes more efficient the process for powering the GRBs jets compared to the standard case, comes from the positive LSB parameter $l>0$ embedded in the background. In other words, rotational energy has no role in sourcing the energy of GRBs.
		The parameter scan performed for $l-a$ within the ranges $0\leq a\leq0.3$ and $-0.5<l\leq0.5$, openly tells us that the model $T=const$ can successfully describe,  in the BG-based black hole, the observed gamma-ray luminosity associated with short and long GRBs ($\sim10^{52-53}$ erg/s), if the LSB parameter falls down in the range $-0.1<l\leq0.3$ (see Fig.~\ref{Con1}). This result is independent of the value of the spin parameter of the black hole. In general, in the model with $T=const$, for not producing GRB with energy higher than the observed one (for short and long cases), one has to set $l \leq0.3$. 

	\subsection{Temperature gradient model}
	\label{Bumblebee - Temperature gradient}
	
\begin{figure}[t]
		\centering
		\includegraphics[scale=0.7]{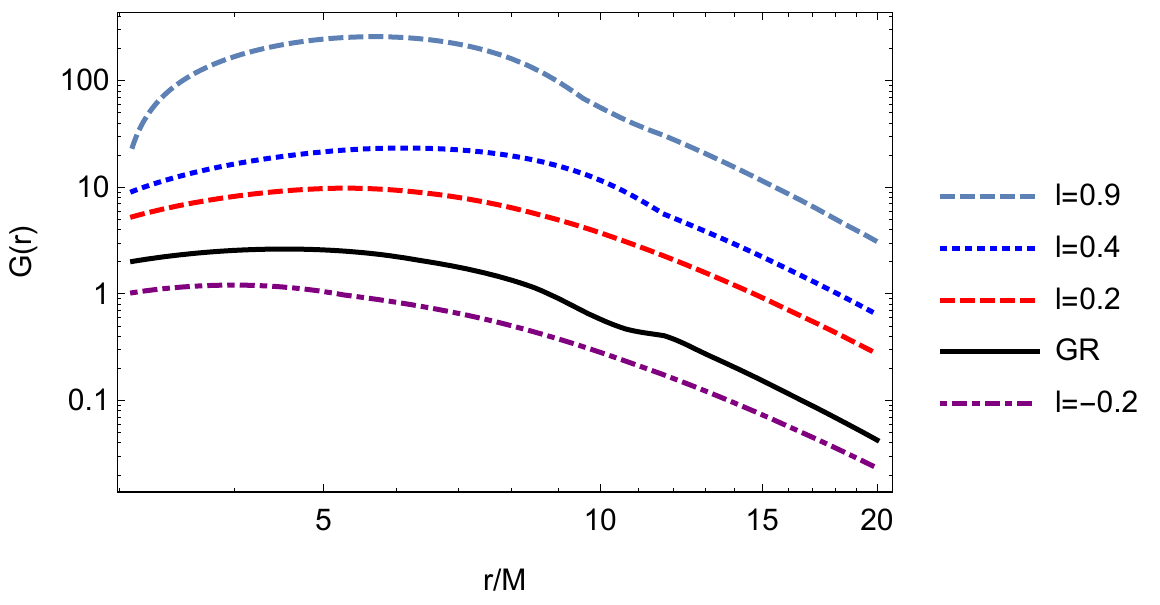}
		\includegraphics[scale=0.7]{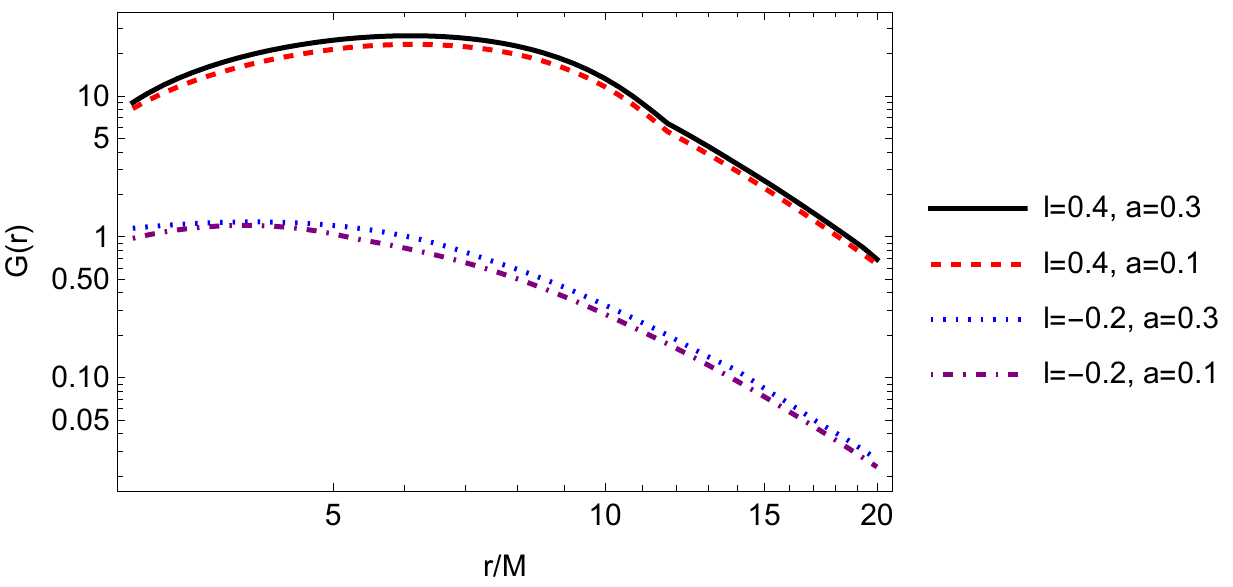}
		\caption{Plot of $G(r)$ against $r/M$ for a disk with a temperature $T\propto 2R_{\mathrm{ph}}/r$. \textbf{Upper graph:}
			Different values of $l$ with fixed value $0.5$ of the spin parameter $a$. \textbf{Lower graph:} Different values of the spin parameter $a$ with $l<0$ and $l>0$.}
		\label{Gr-Tvar}
	\end{figure}

	In the case of a gradient of temperature, the function $G(r)$ is again calculated by using Eq.~(\ref{F(r)}), and taking into account that the temperature varies along with  $r$ ($T_{\rm eff}\propto r^{-1}$), as well as along with  $\theta_{\nu(\bar{\nu})}$. In the upper graph of Fig.~\ref{Gr-Tvar}, we show $G(r)$ vs $r/M$ for different values of the LSB parameter $l$ and for a given value of the slowly rotating parameter of the black hole. Similar to the isothermal disk model, we see here, by moving from $l<0$ to $l>0$, an evident enhancement of the EDR induced by the bumblebee metric as compared to General Relativity ($l=0$). 
		Compared to the upper graph in Fig. \ref{Gr}, one finds that the total energy deposited is lesser than that one expected from the isothermal model. In the lower graph of Fig.~\ref{Gr-Tvar}, we probe, on a case-by-case basis, the role of spin. Unlike the former model, the spin parameter $a$ has a mild effect on the behavior of $G(r)$ related to $l<0$ and $l>0$, so that the EDR for the high value that satisfies the condition $a^2\ll1$, is a bit bigger than lower values.
	\begin{figure}[t]
		\centering
		\includegraphics[scale=0.8]{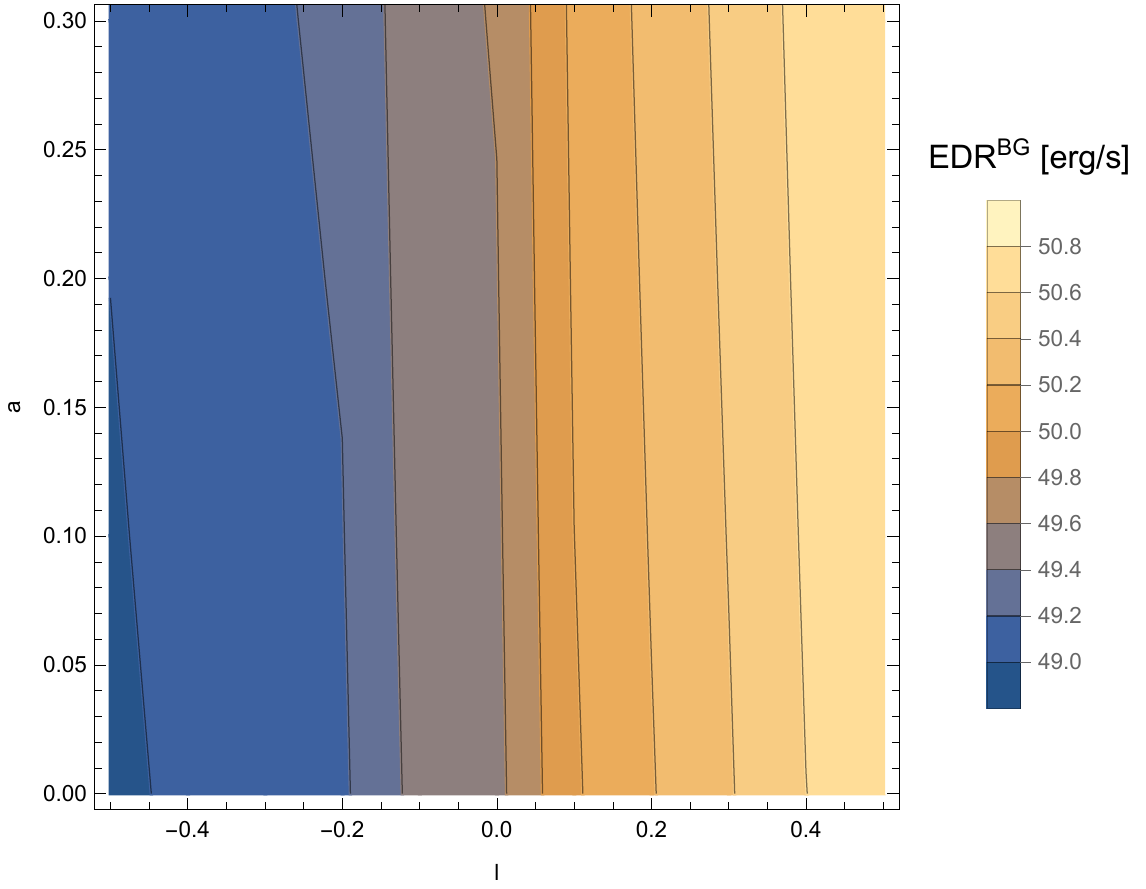}
		\caption{The same contour plot \ref{Con1} with the color scale written in the logarithmic form $\log 10^{n}$ (the range of $n$ is from values lesser than $49$ to beyond $50$), this time for the temperature gradient disk model.}
		\label{Con2}
	\end{figure}
	This mild dependency on the spin parameter $a$ can be traced through the contour plot showing the EDR$^{BG}$ for the temperature gradient model (Fig.~\ref{Con2}). Similar to the case of an isothermal disk, here we also see that the presence of $l>0$ induces an enhancement in EDR. Differently from the case $T=const$, and ignoring the mild dependency on $a$, it is clear that the neutrino annihilation in the environment of the Kerr black hole derived from the BG metric with the accretion disk profile $T\propto r^{-1}$, can only explain ultra-long GRBs ($\sim10^{49-50}$erg/s), if $-0.4\leq l\leq0.1$ (see Fig. \ref{Con2}).
		As a consequence, this model of accretion disk profile is, by covering both negative and positive values of LSB parameter $l$, a suitable candidate for the description of luminosity measured of the ultra-long GRBs jets. 

	\section{Conclusion}
	\label{Conclusion}
	
	It is not clear yet what mechanism conclusively is responsible for launching the gamma-ray bursts (GRBs) jets. The central engine for powering these high energetic jets is usually found in two well-known models: magneto-hydrodynamical and neutrino–antineutrino annihilation ($\nu {\bar \nu}\to e^+ e^-$). These two models, in essence, have been proposed for the energy extraction from a composite system of black hole/accretion disk.
	Concerning the latter (as the mechanism we have considered in this paper), in the case in which the condition $\dot{M}\sim (0.1-1) M_{\odot}$ s$^{-1}$ for the accretion rate of a black hole is satisfied, as well as high enough temperature for the disk, it is expected that the disk can play the role of an efficient neutrino emitter. In this way, the EDR arising from neutrino–antineutrino annihilation at the jet can justify the energetic bursts. In other words, by releasing enormous energy into $e^+ e^-$ pairs by the EDR and subsequently the annihilation process, it is supplied energy to power high energetic photons.
	
	In this paper, we have studied the GRBs jets generated by neutrino pair annihilation for the case in which this process occurs in a slowly rotating Kerr black hole metric (near the rotational axis) modified by the Lorentz symmetry breaking (LSB) parameter $l$. The latter comes from the non-zero VEV of the background bumblebee field $B_\mu$. By employing two idealized models of the accretion disk, one the temperature $T_{\rm eff}=const$, and the other a gradient of temperature for which $T_{\rm eff}\sim r^{-1}$, we have shown that in the presence of the LSB parameter $l>0$, there is an enhancement of the EDR associated with the neutrino-antineutrino annihilation into electron-positron pairs, powering in such a way the GRBs jets. Concerning the first model of the accretion disk, the embedding $l>0$ into spacetime results in an improved situation, compared to the standard slowly rotating Kerr black hole, which is compatible with the observed gamma-ray luminosity associated with the short and long GRBs jets. In this regard, by doing a contour plot analysis of EDR into parameter space $l-a$, we have extracted the upper bound $l\leq 0.3$ for the LSB parameter. The same analysis for the second model has shown that, in the chosen range $-0.4\leq l\leq0.1$, the neutrino EDR can justify the observed luminosity of ultra-long GRBs. Moreover, we want to point out that in both models, the allowed range of $l$ is mostly independent of the spin parameter $a$. In other words, the additional contribution to EDR of GRBs jets arising from neutrino–antineutrino annihilation around a BG-based slowly rotating Kerr black hole, merely comes from the bumblebee vector field embedded in the spacetime background. This can be considered worthwhile since addresses the constructive role of fundamental modifications in explaining GRBs observed in the universe.
	
	As a final comment, a follow-up of the present work would be, indeed, to implement a realist simulation for the disk temperature in the metric (\ref{metricg}) by firstly constructing quasi-stationary disk models as in Ref.~\cite{Popham:1998ab} and then releasing a self-consistent multi-dimensional simulation model.

	\begin{acknowledgments}
		M.Kh. thanks Shiraz University Research Council.	The work of G.L. and L.M. is supported by the Italian Istituto Nazionale di Fisica Nucleare (INFN) through the ``QGSKY'' project and by Ministero dell'Istruzione, Universit\`a e Ricerca (MIUR).
		The computational work has been executed on the IT resources of the ReCaS-Bari data center, which have been made available by two projects financed by the MIUR (Italian Ministry for Education, University and Re-search) in the "PON Ricerca e Competitività 2007-2013" Program: ReCaS (Azione I - Interventi di rafforzamento strutturale, PONa3\_00052, Avviso 254/Ric) and PRISMA (Asse II - Sostegno all'innovazione, PON04a2A)
	\end{acknowledgments}
	
	\appendix
	
	\section{Formulas}
	
	The expression of $F(r, T_0)$ entering Eq. (\ref{trajectory}) is given by
	\begin{widetext}
		\begin{eqnarray}
			F(r,T_0) &=& \frac{2\pi^2}{T^9_{\mathrm{eff}}(2R_{ph})}g^4_{11}(r,0)\Bigg(
			2\int_{\theta_m}^{\theta_M}d\theta_{\nu}T_0^5(\theta_{\nu})\sin\theta_{\nu}\int_{\theta_m}^{\theta_M}d\theta_{\bar{\nu}}T_0^4(\theta_{\bar{\nu}})\sin\theta_{\bar{\nu}} \label{F(r)}
			\\
			&+ &
			\int_{\theta_m}^{\theta_M}d\theta_{\nu}T_0^5(\theta_{\nu})\sin^3\theta_{\nu}\int_{\theta_m}^{\theta_M}d\theta_{\bar{\nu}}T_0^4(\theta_{\bar{\nu}})\sin^3\theta_{\bar{\nu}} \nonumber \\
			& + &
			2\int_{\theta_m}^{\theta_M}d\theta_{\nu}T_0^5(\theta_{\nu})\cos^2\theta_{\nu}\sin\theta_{\nu}\int_{\theta_m}^{\theta_M}d\theta_{\bar{\nu}}T_0^4(\theta_{\bar{\nu}})\cos^2\theta_{\bar{\nu}}\sin\theta_{\bar{\nu}} \nonumber \\
			& - &
			4\int_{\theta_m}^{\theta_M}d\theta_{\nu}T_0^5(\theta_{\nu})\cos\theta_{\nu}\sin\theta_{\nu}\int_{\theta_m}^{\theta_M}d\theta_{\bar{\nu}}T_0^4(\theta_{\bar{\nu}})\cos\theta_{\bar{\nu}}\sin\theta_{\bar{\nu}}\Bigg) \,. \nonumber
		\end{eqnarray}
	\end{widetext}
	
	
\end{document}